\documentclass[11pt]{article}
\usepackage[final]{graphics}

\newcommand{\slL}{\raise.15ex\hbox{$/$}\kern-.53em\hbox{$L$}}
\newcommand{\slP}{\raise.15ex\hbox{$/$}\kern-.53em\hbox{$P$}}
\newcommand{\slR}{\raise.15ex\hbox{$/$}\kern-.53em\hbox{$R$}}
\newcommand{\slQ}{\raise.15ex\hbox{$/$}\kern-.53em\hbox{$Q$}}
\newcommand{\slK}{\raise.15ex\hbox{$/$}\kern-.53em\hbox{$K$}}
\newcommand{\slD}{\raise.15ex\hbox{$/$}\kern-.7em\hbox{$D$}}
\newcommand{\slSigma}{\raise.15ex\hbox{$/$}\kern-.53em\hbox{$\Sigma$}}
\newcommand{\slcalP}{\raise.15ex\hbox{$/$}\kern-.63em\hbox{$\cal P$}}

\font\tenimbf=cmmib10 at 12pt
\font\sevenimbf=cmmib10 at 7pt
\font\fiveimbf=cmmib10 at 5pt
\newfam\imbf
\textfont\imbf=\tenimbf
\scriptfont\imbf=\sevenimbf
\scriptscriptfont\imbf=\fiveimbf
\def\imb{\fam\imbf\tenimbf}

\begin{document}
\begin{titlepage}
\title{\begin{center}
{\Huge LAPTH}
\end{center}
\vspace{5 mm}
\hrule
\vspace{20mm}
\bf{Ambiguities in the zero momentum limit\\
of the thermal $\pi^o\gamma\gamma$ triangle diagram}}
\author{Fran\c cois~~Gelis}
\maketitle

\begin{center}
Laboratoire de Physique Th\'eorique LAPTH,\\
URA 1436 du CNRS, associ\'ee \`a l'universit\'e de Savoie,\\
BP110, F-74941, Annecy le Vieux Cedex, France
\end{center}

\vskip 1cm

\begin{abstract}
  Modifications of the $\pi^o\to 2\gamma$ decay amplitude by thermal
  effects have already been considered by several authors, leading to
  quite different results. I consider in this paper the triangle
  diagram connecting a neutral pion to two photons in a constituent
  quark model, within the real-time formulation of thermal field
  theory and study the zero external momentum limit of this diagram.
  It appears that this limit is not unique and depends strongly on the
  kinematical configuration of the external particles. This
  non-uniqueness is shown to explain the contradiction between
  existing results. I end with some considerations suggesting that
  this decay amplitude may be significantly modified by the
  resummation of hard thermal loops, due to infrared singularities.
\end{abstract}
\vskip 4mm
\centerline{\hfill hep-ph/9806425 \ \ \ LAPTH--689/98\hglue 2cm}
\vfill
\thispagestyle{empty}
\end{titlepage}

\section{Introduction}
During the past two years, a lot of work has been devoted to the study
of the relationship between the axial anomaly and the $\pi^o\to
2\gamma$ decay rate at finite temperature, most notably by Pisarski,
Tytgat and Trueman
\cite{Pisar8,Pisar9,PisarT3,PisarT4,PisarTT1,PisarTT2}.  The purpose
of this series of papers was to explain the following basic fact: the
coefficient of the axial anomaly is independent of the temperature
while the amplitude for the $\pi^o\to 2\gamma$ decay is modified. The
problem was therefore to explain why the relationship that relates at
zero temperature the pion decay amplitude to the axial anomaly ceased
to be valid in a hot medium.

This work has been initiated by a calculation of the pion decay rate
in a constituent quark model, performed by Pisarski in the imaginary
time formalism \cite{Pisar8,Pisar9}. More precisely, it consists in
the calculation of the triangle diagram connecting the pseudo-scalar
to the two emitted photons, via a quark loop. This diagram is
considered in the limit of vanishing external momenta. The result
found in \cite{Pisar8} is that this diagram is proportional to $m/T^2$
where $m$ is the mass of the quark in the loop and $T$ the temperature
of the heat bath, while the result found at zero temperature for the
same diagram is proportional to $1/m$. The consequence of this result
is that the pion decay rate into two photons vanishes if the chiral
symmetry is restored at high temperature, since $m\to 0$.

The same diagram has been calculated in the real time formalism by
\cite{ContrL1,NicolA1}, and also by Gupta and Nayak (GN in the
following) in \cite{GuptaN1} who studied the zero momentum limit of
this diagram.  GN's result for this diagram in the zero external
momentum limit is proportional to $m/mT$.  The dramatic difference is
the behavior of this decay amplitude as a function of the quark mass,
because this behavior was crucial in Pisarski's calculation
\cite{Pisar8,Pisar9} to derive his conclusion about the pion decay
rate in a hot chirally symmetric phase.

The purpose of the present paper is to reconsider the calculation of
the triangle diagram already studied by Pisarski and Gupta \& Nayak,
in order to explain the discrepancy between the results they found. To
that effect, we perform this calculation in the ``retarded-advanced''
version of the real time formalism, but we stay at a more general
level than \cite{Pisar8,ContrL1,NicolA1,GuptaN1} concerning the
kinematical configuration of the external particles. In particular, we
don't assume that external particles are on-shell. Like
\cite{Pisar8,GuptaN1}, we are interested in the zero momentum limit
for this diagram. We arrive at the conclusion that this discrepancy is
due to the non-uniqueness of the zero momentum limit of the considered
Green's function. It appears indeed that this limit depends on the
kinematical configuration of the external legs and that Pisarski and
GN's calculations correspond to very different configurations, GN's
configuration being the most appropriate for the decay of a pion into
real photons. Then, we come back to Pisarski's result about the pion
decay rate in a hot chiral phase and show that, because of infrared
singularities, it may remain valid in GN's kinematical configuration
despite a different dependence in the mass $m$ if one considers the
correction provided by hard thermal loops.

In section \ref{sec:triangle}, we derive the expression for the
triangle diagram in the retarded-advanced formalism, and its
relationship with the pion decay rate. Then, we prove the existence of
a limit of zero external momentum, in a sense to be made precise
later.

In section \ref{sec:limits}, we first give an expression for the zero
momentum limit showing clearly that this limit is not unique and
depends on the kinematical configuration of the external particles.
The remaining of this section is devoted to the detailed study of this
limit in three particular configurations. The first configuration
studied corresponds to a situation where both of the emitted photons
have zero energy: the zero momentum limit reproduces in this case
Pisarski's result. The second important case is obtained with real
photons and a pion at rest in the frame of the plasma: this case
reproduces GN's result. Finally, a third simple case corresponds to
the decay of a static pion into two static photons.

In section \ref{sec:pions}, we study the implications of the above
results for Pisarski's assertion concerning the pion decay amplitude
in a hot chiral phase. Despite the fact that this assertion seems
incorrect at first sight if one considers the physical situation in
which the photons are real, the interplay of infrared singularities in
this calculation makes the resummation of hard thermal loops
necessary. The consequence of this resummation is to change the
parameter playing the role of an infrared regulator. This has the
effect of making the pion decay amplitude vanish in a hot chirally
symmetric phase, even when one is considering the decay into real
photons.

Technical details are relegated to three appendices. In appendix
\ref{app:translation}, we remind the reader of the potentially
dangerous effect of changing the variables in divergent expressions
since this is of some relevance for our calculation. Appendix
\ref{app:AB} gives the general expression of the functions
$A(K_1,K_2)$ and $B(K_1,K_2)$ that appear in intermediate steps of the
calculations. Finally, appendix \ref{app:integrals} gives some details
about a few integrals that appear in this paper.

\section{Triangle diagram in the ``R/A'' formalism}
\label{sec:triangle}
\subsection{$\pi^o$ decay rate}
The decay rate of pions in a thermal bath is related to the
$\pi^o\pi^o$ retarded self-e\-ner\-gy via the relation
\begin{equation}
{{dN}\over{dtd{\imb x}}}=-{{dq_od^3{\imb q}}\over{(2\pi)^4}}\;
2 e^{q_o/T}n_{_{B}}(q_o)\;{\rm Im}\,\Pi^{^{RA}}(q_o,{\imb q})\; ,
\end{equation}
which gives the number of $\pi^o$ decays per unit time and per unit
volume of the plasma, in the four momentum range $dq_od^3{\imb q}$.
The imaginary part of the $\pi^o\pi^o$ two-point function is a sum
over all the possible cuts through the corresponding diagram, which
means that this formula gives the total decay rate, {\it i.e.} the sum
of the contribution of all the channels.  In order to select a
particular channel, one must look at the appropriate cut.

\begin{figure}[htbp]
  \centerline{ \resizebox*{!}{3.5cm}{\includegraphics{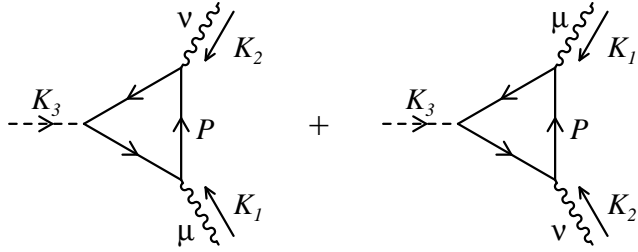}} }
  \caption{\footnotesize{1-loop triangle diagrams contributing to 
      $\pi^o\to \gamma\gamma$.}}
  \label{fig:triangles}
\end{figure}

Like \cite{Pisar8,GuptaN1}, I use a linear sigma model (see
\cite{Koch1} for instance) where the fermion fields are constituent
quarks, in which the mesons are coupled to quark fields as indicated
by the following Lagrangian
\begin{equation}
{\cal L}=i\overline{\Psi}\,\slD\Psi-2g\overline{\Psi}\,(\sigma t_o
+i\mbox{\boldmath$\pi$}\cdot{\imb t}\gamma^5)\Psi\; .
\label{eq:lagrangian}
\end{equation}
I consider two flavors of quarks and $N=3$ colors. The $t$ matrices
are normalized with $t_o=1/2$ and ${\rm Tr}(t_a t_b) =
\delta_{ab}/2$. This coupling is invariant under the chiral symmetry
$SU(2)_{_{L}}\times SU(2)_{_{R}}$. When this symmetry is spontaneously
broken, the $\sigma$ field acquires a non vanishing vacuum expectation
value\footnote{This vacuum expectation value can be identified with
the pion decay constant $f_\pi$ for two flavors at zero
temperature. At nonzero temperature, they differ somehow. Anyway, both
of them vanish when the chiral symmetry is restored.}
$\left<\sigma\right>$, which gives a mass $m=g\left<\sigma\right>$ to
the constituent quarks. In this model, the decay of pions in two
photons appear only in the discontinuity of the three loop
$\pi^o\pi^o$ self-energy.
\begin{figure}[htbp]
  \centerline{ \resizebox*{!}{2.5cm}{\includegraphics{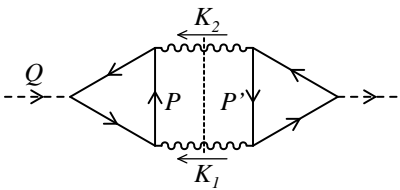}} }
  \caption{\footnotesize{Self-energy of the pseudo-scalar 
      involved in the decay in 2$\gamma$.}}
  \label{fig:self}
\end{figure}
Indeed, each external pseudo-scalars must be connected to a quark
loop, and these two loops must be linked by the two photons. Then,
among all the possible cuts, one must consider the cut that crosses
the photon propagators (see figure \ref{fig:self}). Making use of the
cutting rules for the ``R/A'' formalism \cite{Gelis3}, we find that
the cut depicted on figure \ref{fig:self} contributes:
\begin{eqnarray}
  &&{\rm Im}\,\Pi^{^{RA}}(q_o,{\imb q})=-{1\over
    2}\int{{d^4K_1}\over{(2\pi)^4}} \int{{d^4K_2}\over{(2\pi)^4}} 2\pi
  \epsilon(k_1^o)\delta(K_1^2) 2\pi\epsilon(k_2^o) \delta(K_2^2)
  \nonumber\\
  &&\times (2\pi)^4\delta(Q+K_1+K_2)\;
  \Gamma^{^{ARR}}_{\mu\nu}(Q,K_1,K_2)\Gamma^{^{RAA}}{}^{\mu\nu}(Q,K_1,K_2)\;
  ,
\end{eqnarray}
where $\Gamma_{\mu\nu}^{^{ARR}}(Q,K_1,K_2)$ is the triangle diagram
connecting the pseudo-scalar to two photons. This object will be the
subject of our study from now on ($\Gamma^{^{RAA}}_{\mu\nu}$ is
closely related to the previous one). In fact, two diagrams contribute
to this one-loop 3-point function because of the possibility of
crossing the photons in the final state, as outlined on the figure
\ref{fig:triangles}.
In order to take the two configurations into account, it is sufficient
to calculate in detail the first one, and then add the term obtained
by interchanging the indices $(1,\mu)\leftrightarrow(2,\nu)$.

\subsection{Matrix element}
Let us first give the value of the vertex function
$\Gamma^{^{ARR}}_{\mu\nu}$.  Using the Feynman's rules established for
the ``R/A'' formalism (see \cite{AurenB1,EijckKW1}), a straightforward
calculation gives for one flavor of electric charge $e$:
\begin{eqnarray}
  &&\Gamma^{^{ARR}}_{\mu\nu}(K_3,K_1,K_2)=4\,mN\, e^2 g
  \,\epsilon_{\mu\nu\alpha\beta}\,
  k_1^\alpha k_2^\beta \nonumber\\
  &&\times\int {{d^4P}\over{(2\pi)^4}}
  \left\{n_{_{F}}(p^o+k_2^o)S^{^{A}}(P-K_1)S^{^{A}}(P) \;
    {\rm Disc}\,S^{^{R}}(P+K_2)\right.\nonumber\\
  &&\qquad +n_{_{F}}(p^o)S^{^{A}}(P-K_1) S^{^{R}}(P+K_2)\;
  {\rm Disc}\,S^{^{R}}(P)\nonumber\\
  &&\left. \qquad + n_{_{F}}(p^o-k_1^o)S^{^{R}}(P) S^{^{R}}(P+K_2)\;
    {\rm Disc}\, S^{^{R}}(P-K_1)\right\}\nonumber\\
  &&+\;(K_1,\mu)\leftrightarrow(K_2,\nu)\; ,
\label{eq:vertex}
\end{eqnarray}
where $m$ is the mass of the quark running in the
loop\footnote{Because the vertex coupling the pion to the quark loop
is $g\gamma^5$, the result is proportional to the mass of the
quark. If we were in a chirally symmetric model ($m=0$), the Dirac's
trace would be vanishing.}  and $S^{^{R,A}}(P)\equiv i/(P^2-m^2\pm
ip^o 0^+)$ the scalar part of the retarded (advanced) quark
propagator. In the following, we can forget about the retarded or
advanced labels for the denominators.  Indeed, to recover the correct
prescriptions, it is sufficient to perform at the very end of the
calculation the substitutions:
\begin{equation}
k_1^o\to k_1^o+i0^+\; ,
\quad k_2^o\to k_2^o+i0^+\; ,
\quad k_3^o\to k_3^o-2i0^+\; .
\label{eq:analytic}
\end{equation}

It is worth recalling that the discontinuity of the quark propagator
generates a Dirac's delta function
\begin{equation}
{\rm Disc}\,S^{^{R}}(P)=2\pi\,\epsilon(p^o)\,\delta(P^2-m^2)\;
\end{equation}
which enables us to do easily one of the integrations. To be more
definite, it is convenient to use these Dirac's functions to perform
the integration over the variable $p^o$, so that we are left with a
three-dimensional integration:
\begin{eqnarray}
  &&\Gamma^{^{ARR}}_{\mu\nu}(K_3,K_1,K_2)=4\,mN\, e^2 g
  \,\epsilon_{\mu\nu\alpha\beta}\, k_1^\alpha k_2^\beta \;\int
  {{d^3{\imb p}}\over{(2\pi)^3}}
  \sum\limits_{\epsilon=\pm}\nonumber\\
  &&\!\!\!\!\times \left\{{{n_{_{F}}(\omega_{{\imb p}+{\imb
            k}_2})\!-\!\theta(-\epsilon)}\over {2\omega_{{\imb
            p}+{\imb k}_2}}} \left.  {i\over{2P\cdot
          K_2+K_2^2}}\;{i\over{2P\cdot(K_1+K_2)+K_2^2-K_1^2}}
    \right|_{{{p^o+k_2^o=}\atop{\epsilon\omega_{{\imb p}+{\imb
              k}_2}}}} \right.
  \nonumber\\
  &&+{{n_{_{F}}(\omega_{{\imb p}})-\theta(-\epsilon)}\over
    {2\omega_{{\imb p}}}}\;\left.  {i\over{-2P\cdot
        K_1+K_1^2}}\;{i\over{2P\cdot K_2+K_2^2}}
  \right|_{p^o=\epsilon\omega_{{\imb p}}}\nonumber\\
  &&\!+{{n_{_{F}}(\omega_{{\imb p}-{\imb
          k}_1})\!-\!\theta(-\epsilon)}\over {2\omega_{{\imb p}-{\imb
          k}_1}}} \left.\left. \! {i\over{2P\cdot
          K_1-K_1^2}}\;{i\over{2P\cdot(K_1+K_2)+K_2^2-K_1^2}}
    \right|_{{{p^o-k_1^o=}\atop{\epsilon\omega_{{\imb p}-{\imb
              k}_1}}}}\!
  \right\}\nonumber\\
  &&+\;(K_1,\mu)\leftrightarrow(K_2,\nu)\; ,
\label{eq:3vertex}
\end{eqnarray}
where we denote $\omega_{\imb p}\equiv\surd{({\imb p}^2+m^2)}$. This
expression\footnote{The reader who may wonder why we don't replace
  $P+K_2$ by $P$ in the first term and $P-K_1$ by $P$ in the third one
  is referred to appendix \ref{app:translation}.} of the vertex
function will be the basis of further considerations.

\subsection{Existence of a zero external momenta limit}
We are interested now in the zero momentum limit of this vertex
function in order to understand the origin of the discrepancy between
Pisarski's and GN's result. Only two of the three external momenta are
independent ones due to the energy-momentum conservation: therefore we
choose to consider $K_1$ and $K_2$ as independent momenta and replace
everywhere\footnote{From now on, we drop the explicit reference to the
  argument $K_3$ in $\Gamma^{^{ARR}}_{\mu\nu}$.}  $K_3$ by $-K_1-K_2$.
A priori, taking the limit $K_1,K_2\to 0$ is a very intricate task
since we have to take to zero the eight components of these
four-vectors. In order to simplify without reducing significantly the
generality of the result\footnote{The most general case would be a
  situation where the eight components of $K_{1,2}$ are arbitrary
  functions of the parameter $\lambda$, vanishing when $\lambda\to 0$.
  For our purpose, it is sufficient to restrict ourselves to a linear
  dependence in $\lambda$ of these components.}, we will assume that
the size of the eight components is controlled by some scale
$\lambda$, and this parameter will be the only one taken to zero. This
amounts to write:
\begin{equation}
K_1 \equiv \lambda \hat{K}_1\; ,
\quad K_2 \equiv \lambda \hat{K}_2\; ,
\end{equation}
where the components of $\hat{K}_{1,2}$ are fixed and of order unity.
By this substitution, we are lead to considering the limit when
$\lambda\to 0$ of a univariate function $F(\lambda)$, the
$\hat{K}_{1,2}$ playing the role of constant parameters.

We now want to show that the integral appearing in
Eq.~(\ref{eq:3vertex}) has a finite limit when $\lambda\to 0$.  If we
recall Eq.~(\ref{eq:3vertex}), we can see that this integral is the
sum of six terms (three terms, plus the terms obtained in the
symmetrization with respect to the external photons), each term
behaving like $\lambda^{-2}$ in the limit $\lambda\to 0$.  Therefore,
in order to obtain a finite result, we must expand the integrand in
Eq.~(\ref{eq:3vertex}) up to the order $\lambda^{0}$, and show that we
have cancellations among the various terms in order to eliminate the
orders $\lambda^{-2}$ and $\lambda^{-1}$.

The order $\lambda^{-2}$ is easy to obtain, since we can drop the
$\lambda$ dependence in the statistical functions to extract it, which
gives:
\begin{eqnarray}
  &&\Gamma^{^{ARR}}_{\mu\nu}(K_1,K_2)|_{\lambda^{-2}}={{4mN e^2
  g}\over{\lambda^2}} \,\epsilon_{\mu\nu\alpha\beta}\, {k}_1^\alpha
  {k}_2^\beta \int {{d^3{\imb p}}\over{(2\pi)^3}}\!\!
  \sum\limits_{\epsilon=\pm}{{n_{_{F}}(\omega_{{\imb
  p}})-\theta(-\epsilon)}\over {2\omega_{{\imb p}}}}\nonumber\\
  &&\times \left\{\left.  {i\over{2P\cdot
  \hat{K}_2}}{i\over{2P\cdot(\hat{K}_1+\hat{K}_2)}}
  \right|_{p^o=\epsilon\omega_{{\imb p}}}\left.- {i\over{2P\cdot
  \hat{K}_1}}{i\over{2P\cdot \hat{K}_2}}
  \right|_{p^o=\epsilon\omega_{{\imb p}}}\right.\nonumber\\
  &&+\left.\left.  {i\over{2P\cdot
  \hat{K}_1}}{i\over{2P\cdot(\hat{K}_1+\hat{K}_2)}}
  \right|_{p^o=\epsilon\omega_{{\imb p}}}\right\}
  +\;(\hat{K}_1,\mu)\leftrightarrow(\hat{K}_2,\nu)=0\; .
\label{eq:2vertex}
\end{eqnarray}
As we can see, the cancellation of the order $\lambda^{-2}$ is in fact
a consequence of the energy-momentum conservation (it works because we
have replaced $K_3$ by $-K_1-K_2$).

The cancellation of the order $\lambda^{-1}$ is a consequence of the
parity properties in $K_1$ and $K_2$ of the vertex function. Indeed,
looking at Eq.~(\ref{eq:3vertex}), it is rather straightforward to
check the identity:
\begin{equation}
\Gamma^{^{ARR}}_{\mu\nu}(K_1,K_2)=\Gamma^{^{ARR}}_{\mu\nu}(-K_1,-K_2)\; .
\end{equation}
Making use of the variable $\lambda$, it can be rewritten as:
\begin{equation}
\Gamma^{^{ARR}}_{\mu\nu}(\lambda,\hat{K}_1,\hat{K}_2)=
\Gamma^{^{ARR}}_{\mu\nu}(-\lambda,\hat{K}_1,\hat{K}_2)\; .
\end{equation}
In other words, the one-loop vertex function is an even function of
$\lambda$.  This implies automatically that the terms of order
$\lambda^{-1}$ in the Laurent's expansion of the integral are
vanishing. For this cancellation to occur, it is essential to perform
the symmetrization with respect to the external photons.

\noindent
Therefore, if we write:
\begin{equation}
\Gamma^{^{ARR}}_{\mu\nu}(\lambda,\hat{K}_1,\hat{K}_2)
=4mNe^2g\epsilon_{\mu\nu\alpha\beta}k_1^\alpha k_2^\beta
\;\widetilde{\Gamma}^{^{ARR}}_{\mu\nu}(\lambda,\hat{K}_1,\hat{K}_2)\; ,
\end{equation}
then $\lim_{\lambda\to
0}\widetilde{\Gamma}^{^{ARR}}_{\mu\nu}(\lambda,\hat{K}_1,\hat{K}_2)$
is finite.

\section{Non-uniqueness of the limit}
\label{sec:limits}
\subsection{Generalities}
After some tedious expansions\footnote{At this stage, once we have
  proven the existence of the limit $\lambda\to 0$, we can speed up
  the calculations by making use of some computer algebra system like
  Maple for instance.}, we find:
\begin{eqnarray}
  \lim_{\lambda\to 0}
  &&\Gamma^{^{ARR}}_{\mu\nu}(\lambda,\hat{K}_1,\hat{K}_2)
  =4mNe^2g\,\epsilon_{\mu\nu\alpha\beta}\,
  k_1^\alpha k_2^\beta\int{{d^3{\imb p}}\over{(2\pi)^3}}\nonumber\\
  &&\times \left\{{3\over 8}\;{{1-2n_{_{F}}(\omega_{\imb p})}
      \over{\omega_{\imb p}^5}}\nonumber\right.\\
  &&-{A(\hat{K}_1,\hat{K}_2)\over 4}\;{{n_{_{F}}^\prime(\omega_{\imb
        p})} \over{\omega_{\imb p}^4}} \;\prod\limits_{i=1}^{3}
  {1\over{\Big[({\cal P}_+\cdot \hat{K_i}) ({\cal P}_-\cdot
      \hat{K_i})\Big]^2}}
  \nonumber\\
  &&\left.-{B(\hat{K}_1,\hat{K}_2)\over 4}\;{{n_{_{F}}^{\prime\prime}
        (\omega_{\imb p})}\over{\omega_{\imb p}^3}}\;
    \prod\limits_{i=1}^{3} {1\over{({\cal P}_+\cdot \hat{K_i}) ({\cal
          P}_-\cdot \hat{K_i})}}\right\}\; ,
\label{eq:limit}
\end{eqnarray}
where we denote ${\cal P}_\pm\equiv(\omega_{\imb p},\pm{\imb p})$. The
functions $A$ and $B$ are quite intricate; since their detailed
expression is not really helpful here, they have been quoted in the
appendix \ref{app:AB}. The fact that the above expression still
depends on $\hat{K}_1$ and $\hat{K}_2$ means that the value of the
zero momentum limit depends upon the path chosen to reach the point
$K_1=K_2=0$ in momentum space. 

The non-uniqueness of the zero momentum limit in this case should not
be a surprise. Examples of such a phenomenon are well known in thermal
field theory. For instance, the same calculation applied to the
$\Pi_{00}^{^{RA}}$ component of the photon polarization tensor in
massless QED leads to
\begin{equation}
    \lim_{\lambda\to
      0}\,\Pi_{00}^{^{RA}}(\lambda,\hat{K})=4e^2\int{{d^3{\imb
          p}}\over{(2\pi)^3}}\; n^{\prime}_{_{F}}(p) {{({\imb
          p}\cdot{\hat{\imb k}})^2}\over{({\cal
          P}_+\cdot\hat{K})({\cal P}_-\cdot\hat{K})}}\; ,
    \label{eq:limit-2-point}
\end{equation}
which is nothing but the HTL contribution to this function.  Here
also, the residual dependence upon $\hat{K}$ indicates the
non-uniqueness of the limit. In both cases, this remaining dependence
on how the small momentum limit is reached implies that the
corresponding term in an effective Lagrangian is non-local.  

There is though an important difference between Eq.~(\ref{eq:limit})
and the HTL amplitudes.  The hard thermal loop approximation consists
in retaining only two orders in the expansion in powers of $\lambda$
(the lowest order is trivially vanishing due to momentum
conservation). In the case of Eq.~(\ref{eq:limit}), we have combined
two diagrams so that the second order is also vanishing. We therefore
need to calculate the third order of this expansion, and this is why
the functions $A$ and $B$ are much more involved than what is usually
encountered in hard thermal loops (Eq.~(\ref{eq:limit-2-point}) for
instance). As a consequence, one may expect that the effective
$\pi^o\gamma\gamma$ coupling near the critical point exhibits a
nonlocality of a completely different nature\footnote{Let us recall
  that near $T=0$, the nonlocality of the anomalous couplings is found
  to be closely related to that of hard thermal loops
  \cite{PisarT4,Manue1}. More precisely, HTL-like amplitudes are
  encountered in thermal corrections at the order $T^2/f^2_{\pi}$ in a
  low temperature expansion. Near the chiral phase transition, we are
  in the opposite limit $T\gg f_\pi$, and it is likely that new
  nonlocal terms appear.}.

Before going on with some specific kinematical configurations, a
comment is worth concerning the zero temperature limit of
Eq.~(\ref{eq:limit}). Since for $m>0$ we have $\lim_{_{T\to 0}}
n_{_{F}}(\omega_{\imb p})=\lim_{_{T\to 0}}
n_{_{F}}^{\prime}(\omega_{\imb p})=\lim_{_{T\to 0}}
n_{_{F}}^{\prime\prime}(\omega_{\imb p})=0$, the zero temperature limit
is trivial:
\begin{equation}
\lim_{\lambda\to 0, T\to0}
  \Gamma^{^{ARR}}_{\mu\nu}(\lambda,\hat{K}_1,\hat{K}_2)
  =4mNe^2g\,\epsilon_{\mu\nu\alpha\beta}\,
  k_1^\alpha k_2^\beta\int{{d^3{\imb p}}\over{(2\pi)^3}}
 {3\over {8\omega_{\imb p}^5}}\; .
\end{equation}
As one can see, the integral is now totally independent of the
kinematical configuration of the external particles. Therefore, the
fact that the numerical coefficient in front of the zero momentum
limit of this diagram may not be uniquely defined is a purely thermal
effect.

\subsection{Space-like photons}
\label{sec:Pis}
A first possibility is to consider the situation where $k_{1,2}^o=0$
while ${\imb k}_{1,2}\not={\imb 0}$. This corresponds to external
space-like photons. In this particular case, the functions $A$ and $B$
become much simpler:
\begin{eqnarray}
&&A(\hat{K}_1,\hat{K}_2)=-3({\imb p}\cdot\hat{\imb k}_1)^4
({\imb p}\cdot\hat{\imb k}_2)^4
({\imb p}\cdot(\hat{\imb k}_1+\hat{\imb k}_2))^4\nonumber\\
&&B(\hat{K}_1,\hat{K}_2)=-({\imb p}\cdot\hat{\imb k}_1)^2
({\imb p}\cdot\hat{\imb k}_2)^2
({\imb p}\cdot(\hat{\imb k}_1+\hat{\imb k}_2))^2\; .
\end{eqnarray}
Plugging these expressions into Eq.~(\ref{eq:limit}), we find:
\begin{eqnarray}
  \lim_{\lambda\to 0}
  &&\Gamma^{^{ARR}}_{\mu\nu}(\lambda,\hat{K}_1,\hat{K}_2)
  =4mNe^2g\,\epsilon_{\mu\nu\alpha\beta}\,
  k_1^\alpha k_2^\beta\int{{d^3{\imb p}}\over{(2\pi)^3}}\nonumber\\
  &&\times \left\{{3\over 8}\;{{1-2n_{_{F}}(\omega_{\imb p})}
      \over{\omega_{\imb p}^5}} +{3\over
      4}\;{{n_{_{F}}^\prime(\omega_{\imb p})} \over{\omega_{\imb
          p}^4}} -{1\over 4}\;{{n_{_{F}}^{\prime\prime} (\omega_{\imb
          p})}\over{\omega_{\imb p}^3}}\; \right\}\; .
\label{eq:Pis-limit}
\end{eqnarray}
We can perform at this point the analytic continuation of
Eq.~(\ref{eq:analytic}). Since the functions $A(\hat{K}_1,\hat{K}_2)$
and $B(\hat{K}_1,\hat{K}_2)$ exactly cancel the denominators of
Eq.~(\ref{eq:limit}), this analytic continuation does not introduce
any imaginary part in the result. This fact is a consequence of a
result proven by Evans \cite{Evans9}, according to which all the
retarded/advanced Green's functions are equal if the external energies
are set to zero.

The angular integration is trivial here since it just amounts to
multiply the result by $4\pi$. It remains to perform the integral over
$p=||{\imb p}||$. This integral cannot be performed analytically if
$m\not=0$, but we can consider performing an expansion of the result
in powers of $m/T$, assuming $m\ll T$. In fact, replacing $m$ by zero
in the expression inside the brackets, we can see that the integral
over $p$ is infrared-safe without the need of this mass. As a
consequence, the first term of the expansion in powers of $m/T$ is
trivial to extract:
\begin{eqnarray}
  \lim_{\lambda\to 0}
  &&\Gamma^{^{ARR}}_{\mu\nu}(\lambda,\hat{K}_1,\hat{K}_2)
  =4mNe^2g\,\epsilon_{\mu\nu\alpha\beta}\, k_1^\alpha
  k_2^\beta\int\limits_{0}^{+\infty}
  {{dp}\over{(2\pi)^2}}\nonumber\\
  &&\times \left\{{3\over 4}\;{{1-2n_{_{F}}(p)} \over{p^3}} +{3\over
      2}\;{{n_{_{F}}^\prime(p)} \over{p^2}} -{1\over
      2}\;{{n_{_{F}}^{\prime\prime} (p)}\over{p}}\; \right\}
  \left(1+{\cal O}\left({m\over T}\right)\right)\; .
\end{eqnarray}
Integrating by parts in order to get rid of the inverse powers of $p$,
we obtain:
\begin{eqnarray}
  \lim_{\lambda\to 0}
  &&\Gamma^{^{ARR}}_{\mu\nu}(\lambda,\hat{K}_1,\hat{K}_2)
  =-{{mNe^2g}\over{4\pi^2T^2}}\,\epsilon_{\mu\nu\alpha\beta}\, k_1^\alpha
  k_2^\beta\nonumber\\
&&\times\int\limits_{0}^{+\infty}
  {dx}\,\ln(x)\hat{n}_{_{F}}^{\prime\prime\prime}(x)
  \left(1+{\cal O}\left({m\over T}\right)\right)\; ,
\end{eqnarray}
where we denote $\hat{n}_{_{F}}(x)\equiv 1/(\exp(x)+1)$ and $x\equiv p/T$.
Making use of
\begin{equation}
\hat{n}_{_{F}}^{\prime\prime\prime}(x)=6\hat{n}_{_{F}}^4(x)
-12\hat{n}_{_{F}}^3(x)+7\hat{n}_{_{F}}^2(x)-\hat{n}_{_{F}}(x)\; ,
\end{equation}
and of Eq.~(\ref{eq:ln-integral}) in appendix \ref{app:integrals}, we
finally find\footnote{The formula (\ref{eq:ln-integral})
 of appendix \ref{app:integrals}
  naturally leads to the quantities $\zeta(-2)$ and
  $\zeta^{\prime}(-2)$. In order to simplify the result, we use the
  identities $\zeta(-2)=0$ and $\zeta^{\prime}(-2)=-\zeta(3)/4\pi^2$
  (see for instance \cite{HaberW2}).}:
\begin{equation}
  \lim_{\lambda\to 0}
  \Gamma^{^{ARR}}_{\mu\nu}(\lambda,\hat{K}_1,\hat{K}_2)
  ={{7\zeta(3)mNe^2g}\over{16\pi^4T^2}}\,
  \epsilon_{\mu\nu\alpha\beta}\, k_1^\alpha
  k_2^\beta
  \left(1+{\cal O}\left({m\over T}\right)\right)\; ,
\end{equation}
which is equivalent to formula (11) of \cite{Pisar8}. Therefore, we
have shown that Pisarski's result, obtained in the imaginary time
formalism with external momenta set to zero right from the beginning,
corresponds in fact to a zero-momentum limit taken with space-like
external photons. 

This fact can be interpreted as follows: since in the imaginary time
formalism the energy component of four vectors is a discrete quantity,
the only possible way of taking the ``zero momentum limit'' in this
formalism is to first set the external ``energies'' to the discrete
value zero, and then consider the limit of zero three momenta. The
above analysis shows that the limit is unique once the external
energies are set to zero (the dependence on $\hat{\imb k}_{1,2}$ has
disappeared in Eq.~(\ref{eq:Pis-limit})), which implies that the
imaginary time formalism leads to a uniquely defined limit that co\"\i
ncides with the result obtained here with space-like photons.

A remark is worth concerning the paper \cite{BaierDK1} by Baier, Dirks
and Kober, who reproduced the result of \cite{Pisar8} in a somewhat
different framework. Instead of calculating the triangle diagram in a
particular model, they considered the Wess-Zumino-Witten
\cite{WessZ1,Witte2} functional near the chiral symmetry restoration.
Intermediate steps of their work involve the calculation in the
imaginary time formalism of a function where the external momenta are
set to zero. It seems that this technical analogy with \cite{Pisar8}
is the reason of the agreement.  Since the zero momentum limit of the
$\pi^o\gamma\gamma$ triangle is not uniquely defined, a complete
calculation of the Wess-Zumino-Witten Lagrangian near the chirally
symmetric phase should be extremely careful when using the imaginary
time formalism (or avoid it), in order to get the correct nonlocality
for the couplings contained in this functional.

\subsection{Real photons}
Gupta and Nayak choosed to consider the decay of a massive pion at
rest in the frame of the plasma into two real photons. This choice
corresponds to the constraints ${\imb k}_1+{\imb k}_2={\imb 0}$ and
$k_1^o=k_2^o=-||{\imb k}_{1,2}||$, implying some simplifications for
the functions $A$ and $B$:
\begin{eqnarray}
  &&A(\hat{K}_1,\hat{K}_2)=16{\hat{k}_1^o}{}^4{\omega_{\imb p}}^4
[(\omega_{\imb p}\hat{k}_1^o)^2
-({\imb p}\cdot\hat{\imb k}_1)^2]^4\nonumber\\
&&B(\hat{K}_1,\hat{K}_2)=4{\hat{k}_1^o}{}^2{\omega_{\imb p}}^2
({\imb p}\cdot\hat{\imb k}_1)^2[(\omega_{\imb p}\hat{k}_1^o)^2
-({\imb p}\cdot\hat{\imb k}_1)^2]\; ,
\end{eqnarray}
and for the vertex function:
\begin{eqnarray}
  &&\lim_{\lambda\to 0}
  \Gamma^{^{ARR}}_{\mu\nu}(\lambda,\hat{K}_1,\hat{K}_2)
  =4mNe^2g\,\epsilon_{\mu\nu\alpha\beta}\,
  k_1^\alpha k_2^\beta\int{{d^3{\imb p}}\over{(2\pi)^3}}\nonumber\\
  &&\quad\times \left\{{3\over 8}\;{{1-2n_{_{F}}(\omega_{\imb p})}
      \over{\omega_{\imb p}^5}} -{1\over
      4}\;{{n_{_{F}}^\prime(\omega_{\imb p})} \over{\omega_{\imb
          p}^4}} -{1\over 4}\;{{n_{_{F}}^{\prime\prime} (\omega_{\imb
          p})}\over{\omega_{\imb p}^3}} {{({\imb p}\cdot\hat{\imb
          k})^2}\over{\omega_{\imb p}^2- ({\imb p}\cdot\hat{\imb
          k})^2}}\; \right\}\, ,
\label{eq:Gupta-limit}
\end{eqnarray}
where we denote $\hat{\imb k}\equiv {\imb k}_1/||{\imb k}_1||$ the
unit vector in the direction of the emission of the first photon.  The
analytic continuation of Eq.~(\ref{eq:analytic}) generates a term
$\delta(\omega^2_{\imb p}-({\imb p}\cdot\hat{\imb k})^2)$.  Anyway,
since $\omega_{\imb p} > p$, the Green's function
$\Gamma^{^{ARR}}_{\mu\nu}$ remains real\footnote{This result is very
  similar to the situation encountered in the calculation of hard
  thermal loops which have an imaginary part only if some external
  momentum is space-like. Here, it still works for external lines on
  the light-cone thanks to the mass $m$ in the loop.}.

As one can see now, the angular integral is not defined if the quark
mass is vanishing, due to a collinear singularity. This could have
been expected since we are looking at the emission of real
photons. The angular integration gives the expression:
\begin{eqnarray}
  &&\lim_{\lambda\to 0}
  \Gamma^{^{ARR}}_{\mu\nu}(\lambda,\hat{K}_1,\hat{K}_2)
  =4mNe^2g\,\epsilon_{\mu\nu\alpha\beta}\, k_1^\alpha
  k_2^\beta\int\limits_{0}^{+\infty}
  {{p^2dp}\over{(2\pi)^2}}\nonumber\\
  &&\!\!\times \left\{{3\over 4}\;{{1-2n_{_{F}}(\omega_{\imb p})}
      \over{\omega_{\imb p}^5}} -{{n_{_{F}}^\prime(\omega_{\imb p})}
      \over{2\omega_{\imb p}^4}} \right.
  \left.+{{n_{_{F}}^{\prime\prime} (\omega_{\imb
          p})}\over{2\omega_{\imb p}^3}} \left(1-{{\omega_{\imb
            p}}\over{2p}}\ln\!\left( {{\omega_{\imb
              p}+p}\over{\omega_{\imb p}-p}}\right)\!\right)\;
  \!\!\!\right\}\, .
\end{eqnarray}
Besides the potential collinear singularity, another dramatic
difference of this case with respect to the previous one lies in the
infrared behavior of the integral. It is now impossible to take the
limit $m\to 0$ in the expression inside the brackets because the
integral over $p$ would behave like $dp/p^2$ at small $p$. This means
that the expansion of the integral in powers of $m/T$ begins with a
term in $1/mT$, to be compared with the $1/T^2$ of the previous
situation.  Integrating by parts the above equation, we can transform
it into:
\begin{equation}
  \lim_{\lambda\to 0}
  \!\Gamma^{^{ARR}}_{\mu\nu}(\lambda,\hat{K}_1,\hat{K}_2)
  \!=\!4mNe^2g\epsilon_{\mu\nu\alpha\beta} k_1^\alpha
  k_2^\beta\!\!\!\int\limits_{0}^{+\infty}\!\!
  {{pdp}\over{(2\pi)^2}}{{1-2n_{_{F}}(\omega_{\imb p})}
    \over{4\omega_{\imb p}^4}}\ln\left( {{\omega_{\imb
          p}+p}\over{\omega_{\imb p}-p}}\right)\, ,
  \label{eq:GN-limit}
\end{equation}
which is equivalent to the result given by Gupta and Nayak for the
decay of a static pion into two real photons (see formula (2.12) of
\cite{GuptaN1}). The first term of the expansion in powers of $m/T$
is:
\begin{equation}
  \lim_{\lambda\to 0}
 \Gamma^{^{ARR}}_{\mu\nu}(\lambda,\hat{K}_1,\hat{K}_2)
  ={{mNe^2g}\over{8\pi mT}}\,\epsilon_{\mu\nu\alpha\beta}\, k_1^\alpha
  k_2^\beta\left(1+{\cal O}\left({m\over T}\right)\right) \; .
\end{equation}

\subsection{Photons at rest}
Another simple case is the situation where the emitted photons are
both massive and produced at rest in the frame of the plasma (they can
subsequently decay into lepton pairs). Therefore, the kinematical
constraints we must enforce are now ${\imb k}_{1,2}={\imb 0}$ while
$k_{1,2}^o\not= 0$. With these constraints, the functions $A$ and $B$
become trivial:
\begin{equation}
A(\hat{K}_1,\hat{K}_2)=0\; ,
\quad B(\hat{K}_1,\hat{K}_2)=0\; ,
\end{equation}
so that we have:
\begin{equation}
  \lim_{\lambda\to 0}
 \Gamma^{^{ARR}}_{\mu\nu}(\lambda,\hat{K}_1,\hat{K}_2)
  =4mNe^2g\,\epsilon_{\mu\nu\alpha\beta}\, k_1^\alpha
  k_2^\beta\int{{d^3{\imb p}}\over{(2\pi)^3}} {3\over
    8}\;{{1-2n_{_{F}}(\omega_{\imb p})} \over{\omega_{\imb p}^5}}\; .
\end{equation}
Again, the angular integration is trivial and for the remaining
integral on the variable $p$ we can only perform an expansion in
powers of $m/T$. The analytic continuation of Eq.~(\ref{eq:analytic})
has no effect on this result.  Here also, this integral is infrared
divergent if we put $m=0$ in the integrand. As a consequence, the
result of the integration behaves as $1/mT$ instead of $1/T^2$. More
precisely, we have:
\begin{equation}
  \label{eq:static-limit}
  \lim_{\lambda\to 0}
 \Gamma^{^{ARR}}_{\mu\nu}(\lambda,\hat{K}_1,\hat{K}_2)
  ={{3mNe^2g}\over{32\pi mT}}\,\epsilon_{\mu\nu\alpha\beta}\, k_1^\alpha
  k_2^\beta\left(1+{\cal O}\left({m\over T}\right)\right) \; .
\end{equation}

\subsection{Epilogue}
The above particular examples have demonstrated clearly the
non-u\-ni\-que\-ness of the zero momentum limit of the triangle
diagram responsible for the pion decay in two photons. Moreover, the
particularity of the first situation must be emphasized: when one
expands the integral in powers of $m/T$, there is a cancellation of
the terms of order $1/mT$ so that the first non vanishing terms is of
order $1/T^2$. A closer look at the functions $A$ and $B$ in appendix
\ref{app:AB} and at Eq. (\ref{eq:limit}) indicates that the point
where both $\hat{k}_1^o=0$ and $\hat{k}_2^o=0$ is quite exceptional,
because the functions $A$ and $B$ vanish faster at small $p$ when
$\hat{k}_1^o=\hat{k}_2^o=0$, implying that the integral is not
infrared sensitive. When at least one photon has $\hat{k}_i^o\not=0$,
then at least two powers of $p$ are replaced by $m$ in the small $p$
behavior of $A$ and $B$, so that the expansion starts at the order
$1/mT$.  Therefore, generically, the $\pi^o\gamma\gamma$ effective
coupling {\it does not vanish\/} in the limit of chiral symmetry
restoration $m\to 0$.

\section{IR sensitivity of $\pi^o\to 2\gamma$ and hard thermal loops}
\label{sec:pions}
\subsection{Preliminaries}
The behavior of the decay rate of the $\pi^o$ into $2\gamma$ when the
chiral symmetry is restored is closely related to the behavior of the
$\Gamma^{^{ARR}}_{\mu\nu}$ function in the limit where the mass $m$
goes to zero. The above study shows how this behavior depends on the
kinematical configuration of the external photons.  In particular, we
observe that the imaginary time calculation performed with external
momenta set to zero does not correspond to the physical situation
where the emitted photons are real, but rather to a situation where
the photons are both space-like. The fact that the imaginary time
calculation does not correspond to real photons could have been
expected thanks to the absence of any collinear singularity in this
approach.

The problem is now that GN's situation, which seems more physical
because the photons are assumed to be real, leads to a very different
behavior for the triangle diagram at small $m$. Indeed, Pisarski's
result behaves like $m/T^2$ and therefore vanish in the limit of
chiral symmetry restoration. On the contrary, GN's result behaves like
$m/mT$ and therefore tends to a non vanishing constant when we
consider the same limit. The question is therefore: is the conclusion
that the $\pi^o\to 2\gamma$ decay rate vanishes if the chiral symmetry
is restored at finite temperature correct, since it has been derived
using the result for space-like photons ? At first sight, it seems
that this conclusion is erroneous, because it makes more sense to
consider the result established for real photons in this context.

\subsection{Infrared sensitivity and hard thermal loops}
\begin{figure}[ht]
  \centerline{ \resizebox*{!}{3cm}{\includegraphics{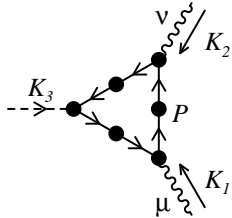}} }
  \caption{\footnotesize{$\pi^o\gamma\gamma$ Green's function 
      at one loop in the HTL expansion.}}
  \label{fig:htl}
\end{figure}

Nevertheless, another aspect of the problem is to be considered, which
may have important consequences in the limit of chiral symmetry
restoration. Indeed, as seen above, the zero momentum limit in the
case of real photons contains a strong infrared divergence, which
gives the factor $1/mT$ (instead of $1/T^2$) once regularized by the
mass $m$. This means that the integral over the loop momentum is
dominated by the soft scale. More accurately, the momentum $p$, of
order $m$, becomes softer and softer as one approaches closer to the
chiral symmetry restoration. Therefore, there is a point when the loop
momentum is soft enough to justify the resummation of hard thermal
loops \cite{BraatP1,FrenkT1,BraatP4,FrenkT2} on the quark propagators,
as outlined on figure \ref{fig:htl}.
Since we have two coupling constants $e$ and $g$ in our model, we can
define two soft scales $eT$ and $gT$. But since the coupling constant
$g$ is related to strong interaction while $e$ comes from
electro-magnetic interactions of the quarks, one may expect that loop
corrections involving the constant $g$ are the dominant ones.
\begin{figure}[ht]
  \centerline{ \resizebox*{!}{3cm}{\includegraphics{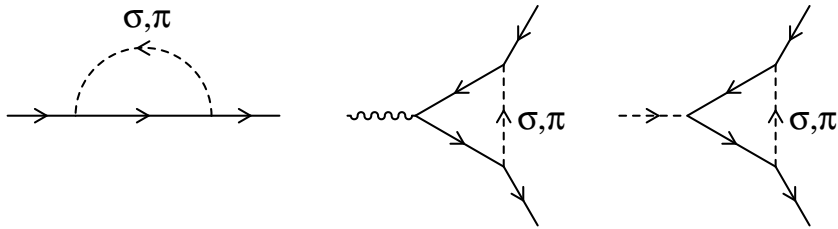}} }
  \caption{\footnotesize{Dominant topologies contributing to the HTLs
  of figure \ref{fig:htl}.}}  \label{fig:g2htl}
\end{figure}
As a consequence, we will consider only loop corrections involving the
$\sigma$ or $\mbox{\boldmath$\pi$}$, as shown in figure
\ref{fig:g2htl}. Looking at the Lagrangian in
Eq.~(\ref{eq:lagrangian}), we see that the coupling of the $\sigma$
field to the quark field is $-ig$ while the coupling of the
$\mbox{\boldmath$\pi$}$ to the quarks is $g\gamma^5$ (I don't write
here the isospin matrices since they appear in such a way that the end
result for the quark self-energy is proportional to the identity in
flavor space). Moreover, in order to derive the contribution of
$\sigma$ and $\mbox{\boldmath$\pi$}$ loops to the HTL correction to
the quark propagator, we can neglect the mass $m$ of the constituent
quarks since $m\ll gT$ when we approach the critical temperature. We
obtain for the retarded self-energy at HTL order:
\begin{eqnarray}
  &&-i\slSigma_{_{RA}}(P)_{_{|\sigma}}=-ig^2
  \int{{d^3{\imb l}}\over{(2\pi)^3}}
  \;{{[n_{_{B}}(l)+n_{_{F}}(l)]}\over{2l\;P\cdot
      \hat{L}}}\; \hat{\slL}\nonumber\\
  &&-i\slSigma_{_{RA}}(P)_{_{|\mbox{\boldmath$\pi$}}}=
  +ig^2\int{{d^3{\imb l}}\over{(2\pi)^3}}\;
  {{[n_{_{B}}(l)+n_{_{F}}(l)]}\over{2l\;P\cdot
      \hat{L}}}\; \gamma^5\hat{\slL}\gamma^5\; ,
\end{eqnarray}
where we denote $\hat{L}\equiv(1,\hat{\imb l})$.  As one can see, the
sum of the above two contributions is precisely equal to the standard
result of QED with $e^2$ replaced by $g^2$. As a consequence, we know
already all the properties of the effective propagator obtained by the
resummation of the above self-energies: this resummation introduces a
cut-off of order $gT$ in loop involving these effective propagators.

For the sake of completeness, we can give expressions for the HTL part
of the above vertices. Starting with the $\gamma q\bar{q}$ vertex, we
find
\begin{equation}
\left.\Gamma_{_{ARR}}^{\gamma
q\bar{q}}(Q,P,-P-Q)\right|^{\mu}_{_{\sigma}}=-ieg^2
\int{{d^3{\imb l}}\over{(2\pi)^3}}
\,{{[n_{_{B}}(l)+n_{_{F}}(l)]\; \hat{\slL}\gamma^\mu\hat{\slL} }
\over{4l\;P\cdot
\hat{L}\;R\cdot\hat{L}}}
\label{eq:vertex-sigma}
\end{equation}
for the contribution of the $\sigma$ field, and 
\begin{equation}
\left.\Gamma_{_{ARR}}^{\gamma
q\bar{q}}(Q,P,-P-Q)\right|^{\mu}_{_{\mbox{\boldmath$\pi$}}}=ieg^2
\int{{d^3{\imb l}}\over{(2\pi)^3}}
\,{{[n_{_{B}}(l)+n_{_{F}}(l)]\;
\gamma^5\hat{\slL}\gamma^\mu\hat{\slL}\gamma^5 }\over{4l\;P\cdot
\hat{L}\;R\cdot\hat{L}}}
\label{eq:vertex-pi}
\end{equation}
for the contribution of the $\mbox{\boldmath$\pi$}$ field, where we
denote $R\equiv P+Q$.  If we add the two contributions, we can see
that it is equal to the QED HTL vertex with two factors of $e$
replaced by $g$. Exactly in the same way, we can obtain the HTL
contribution to the vertex $\pi^o q \bar{q}$. In fact, the result is
obtained by substituting in Eqs.~(\ref{eq:vertex-sigma}) and
(\ref{eq:vertex-pi}) the last power of $e$ by $g$ and the matrix
$\gamma^\mu$ by $i\gamma^5$. Since $\gamma^5$ anti-commutes with the
Dirac's matrices, we see that the product of matrices entering in this
vertex is always proportional to $\slL\slL=L^2=0$. Therefore, the
$\pi^o q\bar{q}$ vertex does not have a HTL contribution at the scale
$gT$ (it may have one at the much smaller scale $eT$ though).

\subsection{Effect of HTLs on $\pi^o\to2\gamma$}
This resummation has the effect of giving a thermal mass $m_{_{T}}$ to
the quark, and this thermal mass remains constant\footnote{Assuming a
  second order phase transition, we expect the coupling constant $g$
  to depend only logarithmically upon temperature, while the vacuum
  expectation value of the sigma field, responsible for the mass $m$
  of the constituent quarks, vanishes as a power of $T-T_c$ at the
  critical temperature $T_c$.} as we approach the point of chiral
symmetry restoration ({\it i.e.} $\lim_{_{T\to T_c}}m_{_{T}}\sim
gT_c$), while the constituent quark mass $m$ goes to zero
($\lim_{_{T\to T_c}}m=0$).  In other words, the thermal mass will
become the relevant infrared regulator as soon as $m\ll m_{_{T}}$
({\it i.e.\/} the relevant infrared regulator is always the biggest
one available).  Moreover, the thermal mass has the property of not
modifying the chiral properties of the propagator which means that if
$m\to 0$, the thermal mass will not change anything to the fact that
the Dirac's trace vanishes. On the basis of these arguments, one can
expect a modification of the $m/mT$ behavior in the case of real
photons.  Indeed, if we track the origin of the various $m$ factors in
this result, we see that the mass $m$ in the numerator comes from the
Dirac's trace, and is closely related to the fact that the chiral
symmetry is broken by the mass $m$. On the contrary, the thermal mass
$m_{_{T}}$ does not break chiral symmetry. Therefore, this factor $m$
at the numerator remains unmodified by the resummation of the thermal
mass. The mass factor in the denominator has a completely different
origin: it comes from the infrared sensitivity of the integration over
the loop momentum. This infrared scale is affected by the resummation
of the thermal mass. As a consequence, we may expect that the mass $m$
is replaced by the thermal mass in the denominator but not in the
numerator, when $m\ll m_{_{T}}$.

\begin{figure}[ht]
  \centerline{ \resizebox*{5cm}{!}{\includegraphics{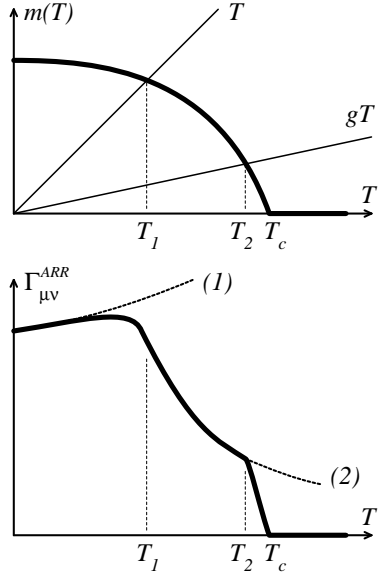}} }
  \caption{\footnotesize{Shape of the temperature dependence of the
  $\pi^o\to2\gamma$ decay amplitude in a constituent quark model. The
  dotted curve $(1)$ is an extrapolation of the $T=0$ result with a
  $T$ dependent quark mass. The dotted curve $(2)$ is the result
  obtained with real external photons when one takes into account
  thermal corrections, but not the resummation of hard thermal
  loops.}}  \label{fig:Tdep}
\end{figure}

Therefore the resummation of the thermal mass $m_{_{T}}$ would lead to
the behavior $m/m_{_{T}}T$ for real photons, near the chiral symmetry
restoration. As a consequence, the result according to which the pion
decay rate for the channel $\pi^o\to 2\gamma$ vanishes in the chiral
phase at finite temperature survives. More precisely, the situation
that emerges from the constituent quark model used in this paper is
summarized on the figure \ref{fig:Tdep}. Assuming that the chiral
symmetry restoration is a second order phase transition, the
constituent quark mass is a function $m=m(T)$ of temperature that
vanishes at a certain critical temperature $T_c$. We can also define
two additional temperature scales $T_1$ and $T_2$ for which
$m(T_1)=T_1$ and $m(T_2)=gT_2$, respectively.

Then, for $T\in[0,T_1]$, the temperature has negligeable effects, and
the decay amplitude is very close to the zero temperature one: it
behaves as $1/m$. When the temperature reaches values of order of
$T_1$, thermal corrections become important, which has the effect of
replacing the factor $1/m^2$ by $1/mT$. As a consequence, the decay
amplitude behaves like $1/T$ for $T\in[T_1,T_2]$. Finally, in the
domain $[T_2,T_c]$, the resummation of thermal masses plays the
dominant role in the regulation of infrared singularities, and the
decay amplitude eventually vanishes for $T=T_c$ since it goes like
$m/gT^2$.

\section{Conclusions}
\label{sec:conslusion}
First, we have seen that the difference between the results of
Pisarski and GN can be interpreted as an effect of the non-uniqueness
of the zero-momentum limit of the triangle diagram at finite
temperature. It appeared also that Pisarski's result, originally
derived in the imaginary-time formalism, corresponds in fact to a
situation where the emitted photons are space-like. The most physical
situation corresponding to the case where the emitted photons are both
real, a superficial analysis tends to invalidate the result according
to which the $\pi^o\to 2\gamma$ decay rate vanishes in a hot chirally
symmetric phase.

Nevertheless, this calculation seems incomplete in the chiral limit
since the loop integral is now sensitive to soft momenta, which means
that the resummation of hard thermal loops may have important effects.
Taking them into account will change the infrared regulator, and
modify GN's result in such a way that it now vanishes in the hot
chiral phase.

I will add also a word of caution concerning the imaginary time
formalism. Since the zero momentum limit of thermal Green's functions
is not unique and depend on the location of the external legs with
respect to the light-cone, this limit cannot be handled correctly in
the imaginary time formalism (there is no light-cone in an Euclidean
formalism). Indeed, in such situations, this formalism gives a number
which corresponds to one particular way of taking the limit, but which
is not necessarily the most appropriate for the problem under study.
Moreover, the information about the nonlocality of the corresponding
effective coupling is lost.

Among the related topics that seem worth studying, I would mention the
case of the box diagram appearing in $\pi^o\sigma\to\gamma\gamma$,
since one can expect here also to have a nonlocal effective coupling
in a hot chirally symmetric phase. Another interesting aspect is to
find a work-around for the limitations of the imaginary time formalism
in the derivation of the anomalous processes near the phase transition
from the Wess-Zumino-Witten Lagrangian.

\section*{Acknowledgments}
First, I would like to thank S.~Gupta for discussions that initiated
this study. I have also to thank the HET group of Brookhaven National
Laboratory for its hospitality and financial support, since an
important part of this work has been done during the period I stayed
in BNL as a summer visitor.  Finally, I would like to thank also
M.~Tytgat and H.~Zaraket for discussions, and P.~Aurenche for many
useful comments on this paper.

\appendix

\section{Side-effects of hazardous translations}
\label{app:translation}
A superficial inspection of the expression of the vertex function
given in Eq.~(\ref{eq:3vertex}) indicates that it may be advantageous
to change the integration variable in the first and third term.
Namely, changing $P+K_2$ into $P$ in the first term and $P-K_1$ into
$P$ in the third term would allow the factorization of a common
statistical weight $(n_{_{F}}(\omega_{\imb
  p})-\theta(-\epsilon))/2\omega_{\imb p}$. Besides this
factorization, the major advantage of such a transformation would be
to eliminate the $K_{1,2}$ dependence ({\it i.e.\/} the $\lambda$
dependence) in this statistical factor: as a consequence, the
subsequent expansions in powers of $\lambda$ would become much
simpler. This technique can be applied in GN's situation where it
leads directly to Eq.~(\ref{eq:GN-limit}), without the need of
performing a cumbersome integration by parts (the reason for that lies
in the fact that no derivatives of $n_{_{F}}$ are generated in this
second approach, because the statistical prefactor does not contain
the expansion parameter $\lambda$). This method works also for the
third case studied in section \ref{sec:limits}.  Nevertheless, we
avoided its use to derive the general limit presented in
Eq.~(\ref{eq:limit}) because this transformation is not always
legitimate.


The reason why the above transformations are sometimes illegitimate
comes in fact from the infrared sector. To be more definite, let me
focus on the situation of Pisarski, since for this one the above
changes of variables are not allowed. The particularity of this
configuration is that $k_1^o=k_2^o=0$. Therefore, the denominators in
Eq.~(\ref{eq:3vertex}) are combinations of $2{\imb p}\cdot{\imb
k}_1-{\imb k}_1^2$ and $2{\imb p}\cdot{\imb k}_2+{\imb k}_2^2$. As a
consequence, the expansion of these factors in powers of $\lambda$
generates powers of ${\imb k}_i^2/({\imb p}\cdot{\imb k}_i)$. A simple
counting shows that the order $\lambda^0$ behaves like $\int dp/p^2$
in the infrared region, even if we keep $m$ strictly positive.
Therefore, each individual term in Eq.~(\ref{eq:3vertex}) is strongly
infrared divergent at $p=0$. Of course, a conspiracy of the three
terms occur in order to cancel this divergence so that the final
result is finite. If one performs different translations on the three
terms, then these finite terms are modified. The correct answer can
only be obtained when the same transformation (or no transformation at
all) is applied to the three terms.

An alternative way to see that this transformation is not legitimate
in Pisarski's situation is as follows. We have seen on the GN's case
that an integration by parts relates the expression obtained without
this transformation to the expression one would obtain by making use
of it. Therefore, let us integrate by parts the result of
Eq.~(\ref{eq:Pis-limit}):
\begin{eqnarray}
  &&\int\limits_{0}^{+\infty} p^2 dp \left\{ 
{3\over 4}{{1-2n_{_{F}}(\omega_{\imb p})}\over{\omega_{\imb p}^5}}
+{3\over 2}{{n_{_{F}}^{\prime}(\omega_{\imb p})}\over{\omega_{\imb p}^4}}
-{1\over 2}{{n_{_{F}}^{\prime\prime}(\omega_{\imb p})}\over{\omega_{\imb p}^3}}
\right\}\nonumber\\
&&=\int\limits_{0}^{+\infty}p^2 dp \left\{
{3\over{4\omega_{\imb p}^5}}
+{1\over 2}{{n_{_{F}}(\omega_{\imb p})}\over{p^4\omega_{\imb p}}}
+{1\over 2}{{n_{_{F}}(m)}\over{\omega_{\imb p}^5}}
\left[
6-{{3\omega_{\imb p}^2}\over{p^2}}-{{\omega_{\imb p}^4}\over{p^4}}
\right]
\right\}\nonumber\\
&&={{1-2n_{_{F}}(m)}\over{4m^2}}
+\int\limits_{0}^{+\infty}dp\;{{n_{_{F}}(\omega_{\imb p})-n_{_{F}}(m)}\over
{2p^2\omega_{\imb p}}}\; .
\end{eqnarray}
As one can see, some terms in $n_{_{F}}(m)$ appear in this integration
by parts, which are absolutely necessary to ensure the infrared
finiteness of the integral. There is no way of transforming this
integral in order to have the temperature dependence only in a factor
$1-2n_{_{F}}(\omega_{\imb p})$, as it would be if the changes of
variables $P+K_2\to P, P-K_1\to P$ were possible\footnote{The
  condition to be able to eliminate the $n_{_{F}}(m)$ terms is
  \begin{equation}
    \lim_{p\to 0}{{\omega_{\imb p}}\over p}{{\partial}\over{\partial p}}
\left[ 
{p\over{\omega_{\imb p}^2}}\int d\Omega_{\imb p}\,
B(\hat{K}_1,\hat{K}_2) \prod_{i=1}^{3} {1\over{({\cal P}_+\cdot \hat{K}_i)
({\cal P}_-\cdot\hat{K}_i)}}
\right]=0\; ,
  \end{equation}
  which is satisfied in GN's case but not in Pisarski's one.  }.

\section{Functions $A(K_1,K_2)$ and $B(K_1,K_2)$}
\label{app:AB}

\begin{eqnarray*} \lefteqn{A(K_1,K_2)=
{k_1^o}^{3}\,{k_2^o}^{5}\,(3\,{k_1^o}+{k_2^o})
\,({k_1^o}\,{{\imb k}_2^2} +{k_2^o}\, {{\imb k}_1^2})\,\omega_p ^{12} }\\
&& + {k_1^o}\,{k_2^o}^{4}\Big[ {k_1^o}\,{k_2^o}^{4}\,{({\imb
    p}\cdot{\imb k}_1)}^{2} \\
&& + {k_1^o}^{2}\,{k_2^o}^{3}\,{({\imb
    p}\cdot{\imb k}_1)}^{2}
+ 2\,{k_1^o}^{2}\,{k_2^o}^{3}\,{({\imb p}\cdot{\imb k}_1)}\,{({\imb p}\cdot{\imb k}_2)}\\
&& + {k_2^o}^{3}\,{{\imb k}_1^2}\,{({\imb p}\cdot{\imb k}_1)}^{2}
+ 10\,{k_1^o}^{3}\,{k_2^o}^{2}\,{({\imb p}\cdot{\imb k}_1)}\,{({\imb p}\cdot{\imb k}_2)}\\
&& - 2\,{k_1^o}^{3}\,{k_2^o}^{2}\,{({\imb p}\cdot{\imb k}_1)}^{2}
- {k_1^o}^{3}\,{k_2^o}^{2}\,{({\imb p}\cdot{\imb k}_2)}^{2} \\
&& + 2\,{k_1^o}\,{k_2^o}^{2}\,{{\imb k}_1^2}\,{({\imb p}\cdot{\imb
    k}_1)}\,{({\imb p}\cdot{\imb k}_2)}
- 3\,{k_1^o}\,{k_2^o}^{2}\,{({\imb p}\cdot{\imb k}_1)}^{2}\,{{\imb k}_2^2} \\
&& + 6\,{k_1^o}\,{k_2^o}^{2}\,{{\imb k}_1^2}\,{({\imb p}\cdot{\imb
    k}_1)}^{2}
+ 4\,{k_1^o}^{2}\,{k_2^o}\,{{\imb k}_1^2}\,{({\imb p}\cdot{\imb k}_1)}\,{({\imb p}\cdot{\imb k}_2)} \\
&& - 10\,{k_1^o}^{2}\,{k_2^o}\,{{\imb k}_2^2}\,{({\imb p}\cdot{\imb
    k}_1)}^{2}
+ 8\,{k_1^o}^{4}\,{k_2^o}\,{({\imb p}\cdot{\imb k}_1)}\,{({\imb p}\cdot{\imb k}_2)} \\
&& - 2\,{k_1^o}^{2}\,{k_2^o}\,{{\imb k}_2^2}\,{({\imb p}\cdot{\imb
    k}_1)}\,{({\imb p}\cdot{\imb k}_2)}
+ 10\,{k_1^o}^{2}\,{k_2^o}\,{{\imb k}_1^2}\,{({\imb p}\cdot{\imb k}_1)}^{2} \\
&& - 3\,{k_1^o}^{4}\,{k_2^o}\,{({\imb p}\cdot{\imb k}_2)}^{2}
- 3\,{k_1^o}^{2}\,{k_2^o}\,{{\imb k}_1^2}\,{({\imb p}\cdot{\imb k}_2)}^{2} \\
&& - 10\,{k_1^o}^{3}\,{{\imb k}_1^2}\,{({\imb p}\cdot{\imb k}_2)}^{2}
+ 5\,{k_1^o}^{3}\,{{\imb k}_2^2}\,{({\imb p}\cdot{\imb k}_2)}^{2}\Big]
\omega_p ^{10}\\
&& + {k_2^o}^{3} \Big[ 4\,{k_1^o}^{3}\,{({\imb p}\cdot{\imb
    k}_1)}\,{{\imb k}_2^2}\,{({\imb p}\cdot{\imb k}_2)}^{3} \\
&& - 8\,{k_1^o}^{3}\,{{\imb k}_1^2}\,{({\imb p}\cdot{\imb k}_1)}\,{({\imb
    p}\cdot{\imb k}_2)}^{3}
- {({\imb p}\cdot{\imb k}_1)}^{4}\,{k_2^o}^{5} \\
&& - 2\,{k_2^o}^{4}\,{k_1^o}\,{({\imb p}\cdot{\imb k}_1)}^{3}\,{({\imb
    p}\cdot{\imb k}_2)}
- 20\,{k_2^o}^{3}\,{k_1^o}^{2}\,{({\imb p}\cdot{\imb k}_1)}^{3}\,{({\imb p}\cdot{\imb k}_2)} \\
&& - 3\,{k_2^o}^{4}\,{k_1^o}\,{({\imb p}\cdot{\imb k}_1)}^{4}
- 26\,{k_2^o}\,{k_1^o}^{4}\,{({\imb p}\cdot{\imb k}_2)}^{3}\,{({\imb p}\cdot{\imb k}_1)} \\
&& - 9\,{k_2^o}\,{k_1^o}^{4}\,{({\imb p}\cdot{\imb k}_2)}^{4}
+ 7\,{k_2^o}^{2}\,{k_1^o}^{3}\,{({\imb p}\cdot{\imb k}_1)}^{2}\,{({\imb p}\cdot{\imb k}_2)}^{2} \\
&& - 36\,{k_2^o}^{2}\,{k_1^o}^{3}\,{({\imb p}\cdot{\imb
    k}_1)}^{3}\,{({\imb p}\cdot{\imb k}_2)}
+ 3\,{k_2^o}\,{({\imb p}\cdot{\imb k}_2)}^{2}\,{k_1^o}^{4}\,{({\imb p}\cdot{\imb k}_1)}^{2} \\
&& - 12\,{k_2^o}^{3}\,{k_1^o}^{2}\,{({\imb p}\cdot{\imb k}_1)}^{4}
- 18\,{k_2^o}^{2}\,{({\imb p}\cdot{\imb k}_1)}^{4}\,{k_1^o}^{3} \\
&& + {k_2^o}^{3}\,{k_1^o}^{2}\,{({\imb p}\cdot{\imb
    k}_1)}^{2}\,{({\imb p}\cdot{\imb k}_2)}^{2}
- 4\,{k_2^o}^{2}\,{k_1^o}^{3}\,{({\imb p}\cdot{\imb k}_1)}\,{({\imb p}\cdot{\imb k}_2)}^{3} \\
&& - 12\,{k_1^o}^{3}\,{({\imb p}\cdot{\imb k}_1)}^{2}\,{{\imb
    k}_2^2}\,{({\imb p}\cdot{\imb k}_2)}^{2}
+ 3\,{k_1^o}^{3}\,{{\imb k}_1^2}\,{({\imb p}\cdot{\imb k}_2)}^{4} \\
&& - 4\,{k_2^o}^{2}\,{k_1^o}\,{{\imb k}_1^2}\,{({\imb p}\cdot{\imb
    k}_1)}^{3}\,{({\imb p}\cdot{\imb k}_2)}
+ 2\,{k_2^o}^{2}\,{k_1^o}\,{({\imb p}\cdot{\imb k}_1)}^{3}\,{{\imb k}_2^2}\,{({\imb p}\cdot{\imb k}_2)} \\
&& + 7\,{k_2^o}^{2}\,{k_1^o}\,{({\imb p}\cdot{\imb k}_1)}^{4}\,{{\imb
    k}_2^2}
- 5\,{k_2^o}\,{k_1^o}^{2}\,{{\imb k}_1^2}\,{({\imb p}\cdot{\imb k}_1)}^{4} \\
&& - 3\,{k_2^o}^{2}\,{k_1^o}\,{{\imb k}_1^2}\,{({\imb p}\cdot{\imb
    k}_1)}^{2}\,{({\imb p}\cdot{\imb k}_2)}^{2}
- 5\,{k_2^o}^{2}\,{{\imb k}_1^2}\,{k_1^o}\,{({\imb p}\cdot{\imb k}_1)}^{4} \\
&& - {k_2^o}^{3}\,{{\imb k}_1^2}\,{({\imb p}\cdot{\imb k}_1)}^{4}
+ 2\,{k_2^o}^{3}\,{({\imb p}\cdot{\imb k}_1)}^{4}\,{{\imb k}_2^2} \\
&& + 7\,{k_2^o}\,{({\imb p}\cdot{\imb k}_1)}^{4}\,{k_1^o}^{2}\,{{\imb
    k}_2^2}
- 6\,{k_2^o}\,{k_1^o}^{2}\,{{\imb k}_1^2}\,{({\imb p}\cdot{\imb k}_1)}\,{({\imb p}\cdot{\imb k}_2)}^{3} \\
&& - 9\,{k_2^o}\,{k_1^o}^{2}\,{({\imb p}\cdot{\imb k}_1)}^{2}\,{{\imb
    k}_2^2}\,{({\imb p}\cdot{\imb k}_2)}^{2}
- 4\,{k_2^o}\,{{\imb k}_1^2}\,{k_1^o}^{2}\,{({\imb p}\cdot{\imb k}_1)}^{3}\,{({\imb p}\cdot{\imb k}_2)} \\
&& - 4\,{k_2^o}\,{k_1^o}^{2}\,{({\imb p}\cdot{\imb k}_1)}^{3}\,{{\imb
    k}_2^2}\,{({\imb p}\cdot{\imb k}_2)}
- 12\,{k_2^o}\,{k_1^o}^{2}\,{{\imb k}_1^2}\,{({\imb p}\cdot{\imb k}_1)}^{2}\,{({\imb p}\cdot{\imb k}_2)}^{2}\Big] \omega_p ^{8}\\
&& + {({\imb p}\cdot{\imb k}_1)}\,{k_2^o}^{2} \Big[
6\,{k_2^o}\,{({\imb p}\cdot{\imb k}_1)}^{3}\,{{\imb k}_1^2}\,{k_1^o}\,{({\imb p}\cdot{\imb k}_2)}^{2} \\
&& + 3\,{k_2^o}\,{({\imb p}\cdot{\imb k}_1)}\,{k_1^o}\, {{\imb
    k}_1^2}\,{({\imb p}\cdot{\imb k}_2)}^{4} - 4\,{k_2^o}\,{({\imb
    p}\cdot{\imb k}_1)}^{2
  }\,{k_1^o}\,{{\imb k}_2^2}\,{({\imb p}\cdot{\imb k}_2)}^{3} \\
& & \mbox{} + 6\,{({\imb p}\cdot{\imb k}_1)}^{5}\,{k_2^o}^{4} + 8\,
{({\imb p}\cdot{\imb k}_1)}^{4}\,{({\imb p}\cdot{\imb
    k}_2)}\,{k_2^o}^{4} \\
&& + 2\,{({\imb p}\cdot{\imb k}_1)
  }^{3}\,{k_1^o}\,{({\imb p}\cdot{\imb k}_2)}^{2}\,{k_2^o}^{3} \\
& & \mbox{} + 17\,{({\imb p}\cdot{\imb k}_1)}\,{k_2^o}^{2}\,{k_1^o}^{
  2}\,{({\imb p}\cdot{\imb k}_2)}^{4} + 20\,{({\imb p}\cdot{\imb
    k}_1)}^{4}\,{k_2^o}^{3}\,
{({\imb p}\cdot{\imb k}_2)}\,{k_1^o} \\
& & \mbox{} + 4\,{({\imb p}\cdot{\imb k}_1)}^{2}\,{k_2^o}^{3}\,{k_1^o
  }\,{({\imb p}\cdot{\imb k}_2)}^{3} + 56\,{({\imb p}\cdot{\imb
    k}_1)}^{2}\,{k_2^o}^{2}\,
{k_1^o}^{2}\,{({\imb p}\cdot{\imb k}_2)}^{3} \\
& & \mbox{} + 13\,{({\imb p}\cdot{\imb k}_1)}^{5}\,{k_1^o}\,{k_2^o}^{
  3} + 13\,{k_2^o}^{2}\,{({\imb p}\cdot{\imb k}_1)}^{5}\,{k_1^o}^{2} \\
& & \mbox{} + 34\,{k_2^o}^{2}\,{({\imb p}\cdot{\imb k}_1)}^{4}\,{
  ({\imb p}\cdot{\imb k}_2)}\,{k_1^o}^{2} + 2\,{k_2^o}\,{k_1^o}^{3}\,
{({\imb p}\cdot{\imb k}_2)}^{5} \\
& & \mbox{} + 31\,{k_2^o}\,{({\imb p}\cdot{\imb k}_1)}\,{k_1^o}^{3}\,
{({\imb p}\cdot{\imb k}_2)}^{4} + 40\,{k_2^o}\,{({\imb p}\cdot{\imb
    k}_1)}^{2}\,{
  k_1^o}^{3}\,{({\imb p}\cdot{\imb k}_2)}^{3} \\
& & \mbox{} + 16\,{({\imb p}\cdot{\imb k}_1)}^{2}\,{k_1^o}^{2}\,{
  {\imb k}_2^2}\,{({\imb p}\cdot{\imb k}_2)}^{3} +
6\,{k_1^o}^{2}\,{{\imb k}_1^2}\,
{({\imb p}\cdot{\imb k}_2)}^{5} \\
& & \mbox{} + 42\,{k_2^o}^{2}\,{({\imb p}\cdot{\imb k}_1)}^{3}\,{
  ({\imb p}\cdot{\imb k}_2)}^{2}\,{k_1^o}^{2} + 14\,{({\imb
    p}\cdot{\imb k}_1)}\,{k_1^o}^{2}\,
{{\imb k}_1^2}\,{({\imb p}\cdot{\imb k}_2)}^{4} \\
& & \mbox{} + 11\,{({\imb p}\cdot{\imb k}_1)}\,{k_1^o}^{2}\,{({\imb
    p}\cdot{\imb k}_2)}^{ 4}\,{{\imb k}_2^2} - {k_2^o}^{2}\,{({\imb
    p}\cdot{\imb k}_1)}^{5}\,{
  {\imb k}_1^2} \\
& & \mbox{} - 2\,{k_2^o}^{2}\,{({\imb p}\cdot{\imb k}_1)}^{3}\,{{\imb
    k}_1^2 }\,{({\imb p}\cdot{\imb k}_2)}^{2} +
4\,{k_2^o}^{2}\,{({\imb p}\cdot{\imb k}_1)}^{3}\,
{{\imb k}_2^2}\,{({\imb p}\cdot{\imb k}_2)}^{2} \\
& & \mbox{} + 4\,{k_2^o}^{2}\,{({\imb p}\cdot{\imb k}_1)}^{4}\,{{\imb
    k}_2^2 }\,{({\imb p}\cdot{\imb k}_2)} - {k_2^o}\,{({\imb
    p}\cdot{\imb k}_1)}^{5}\,{{\imb k}_2^2}\,
{k_1^o} \\
& & \mbox{} + 4\,{k_2^o}\,{({\imb p}\cdot{\imb k}_1)}^{4}\,{{\imb
    k}_2^2}\, {({\imb p}\cdot{\imb k}_2)}\,{k_1^o} -
2\,{k_2^o}\,{({\imb p}\cdot{\imb k}_1)}^{4}\,
{{\imb k}_1^2}\,{({\imb p}\cdot{\imb k}_2)}\,{k_1^o} \\
& & \mbox{} - 4\,{k_2^o}^{2}\,{({\imb p}\cdot{\imb k}_1)}^{4}\,{{\imb
    k}_1^2 }\,{({\imb p}\cdot{\imb k}_2)} + 2\,{k_2^o}\,{({\imb
    p}\cdot{\imb k}_1)}^{3}\,{k_1^o}
\,{{\imb k}_2^2}\,{({\imb p}\cdot{\imb k}_2)}^{2} \\
& & \mbox{} + 8\,{k_2^o}\,{({\imb p}\cdot{\imb k}_1)}^{2}\,{k_1^o}\,
{{\imb k}_1^2}\,{({\imb p}\cdot{\imb k}_2)}^{3}\Big]\omega_p ^{6}\\
&& - {({\imb p}\cdot{\imb k}_1)} ^{2}\,{k_2^o}\Big[
10\,{k_2^o}^{3}\,{({\imb p}\cdot{\imb k}_1)}^{5}\,{({\imb p}\cdot{\imb
    k}_2)} + 22\,
{k_2^o}^{3}\,{({\imb p}\cdot{\imb k}_1)}^{4}\,{({\imb p}\cdot{\imb k}_2)}^{2} \\
& & \mbox{} + 24\,{k_2^o}^{3}\,{({\imb p}\cdot{\imb k}_1)}^{3}\,{
  ({\imb p}\cdot{\imb k}_2)}^{3} + 3\,{k_2^o}^{3}\,{({\imb
    p}\cdot{\imb k}_1)}^{6} + 8\,{k_2^o
  }^{3}\,{({\imb p}\cdot{\imb k}_1)}^{2}\,{({\imb p}\cdot{\imb k}_2)}^{4} \\
& & \mbox{} + 23\,{k_2^o}^{2}\,{({\imb p}\cdot{\imb k}_1)}^{4}\,{
  ({\imb p}\cdot{\imb k}_2)}^{2}\,{k_1^o} + 44\,{k_2^o}^{2}\,{({\imb
    p}\cdot{\imb k}_1)}^{3}\,
{k_1^o}\,{({\imb p}\cdot{\imb k}_2)}^{3} \\
& & \mbox{} + 2\,{k_2^o}^{2}\,{({\imb p}\cdot{\imb k}_1)}\,{k_1^o}\,
{({\imb p}\cdot{\imb k}_2)}^{5} + 19\,{k_2^o}^{2}\,{({\imb
    p}\cdot{\imb k}_1)}^{2}\,
{k_1^o}\,{({\imb p}\cdot{\imb k}_2)}^{4} \\
& & \mbox{} + 12\,{k_2^o}\,{({\imb p}\cdot{\imb k}_1)}^{3}\,{{\imb
    k}_2^2}\, {({\imb p}\cdot{\imb k}_2)}^{3} - 4\,{k_2^o}\,{({\imb
    p}\cdot{\imb k}_1)}^{3}\,{{\imb k}_1^2
  }\,{({\imb p}\cdot{\imb k}_2)}^{3} \\
& & \mbox{} + 25\,{k_2^o}\,{k_1^o}^{2}\,{({\imb p}\cdot{\imb k}_2)}^{
  6} + 76\,{k_2^o}\,{({\imb p}\cdot{\imb k}_1)}\,{k_1^o}^{2}\,{
  ({\imb p}\cdot{\imb k}_2)}^{5} \\
& & \mbox{} + 2\,{k_2^o}\,{({\imb p}\cdot{\imb k}_1)}^{5}\,{{\imb
    k}_2^2}\, {({\imb p}\cdot{\imb k}_2)} - 3\,{k_2^o}\,{({\imb
    p}\cdot{\imb k}_1)}^{2}\,{{\imb k}_1^2}\,
{({\imb p}\cdot{\imb k}_2)}^{4} \\
& & \mbox{} + 8\,{k_2^o}\,{({\imb p}\cdot{\imb k}_1)}^{4}\,{({\imb
    p}\cdot{\imb k}_2)}^{2 }\,{{\imb k}_2^2} - {k_2^o}\,{({\imb
    p}\cdot{\imb k}_1)}^{4}\,{({\imb p}\cdot{\imb k}_2)}^{2
  }\,{{\imb k}_1^2} \\
&& + 48\,{k_2^o}\,{({\imb p}\cdot{\imb k}_1)}^{2}\,{({\imb
    p}\cdot{\imb k}_2)}^{ 4}\,{k_1^o}^{2} + 6\,{k_2^o}\,{({\imb
    p}\cdot{\imb k}_1)}^{2}\,
{{\imb k}_2^2}\,{({\imb p}\cdot{\imb k}_2)}^{4} \\
&& + 4\,{({\imb p}\cdot{\imb k}_1)}\,{k_1^o}\,{{\imb k}_1^2}\, {({\imb
    p}\cdot{\imb k}_2)}^{5} + {k_1^o}\,{({\imb p}\cdot{\imb
    k}_2)}^{6}\,{{\imb k}_1^2}
\\
&& + {({\imb p}\cdot{\imb k}_1)}^{2}\,{k_1^o}\,{{\imb k}_2^2}\,
{({\imb p}\cdot{\imb k}_2)}^{4} - 2\,{({\imb p}\cdot{\imb
    k}_1)}\,{k_1^o}\,{({\imb p}\cdot{\imb k}_2)}^{5
  }\,{{\imb k}_2^2}\Big]\omega_p ^{4}\\
&& + 6 ({({\imb p}\cdot{\imb k}_2)} + {({\imb p}\cdot{\imb
    k}_1)})\,{k_2^o} \,{({\imb p}\cdot{\imb k}_1)}^{4} \,{({\imb
    p}\cdot{\imb k}_2)}^{2}\Big[{k_2^o}\,{({\imb p}\cdot{\imb
    k}_1)}^{3} \\
&& + 3\,{
  ({\imb p}\cdot{\imb k}_1)}^{2}\,{k_2^o}\,{({\imb p}\cdot{\imb k}_2)} \\
& & \mbox{} + 4\,{k_2^o}\,{({\imb p}\cdot{\imb k}_1)}\,{({\imb
    p}\cdot{\imb k}_2)}^{2} + 2\,{k_2^o}\,{({\imb p}\cdot{\imb
    k}_2)}^{3} + 2\,{k_1^o}\,{({\imb p}\cdot{\imb k}_2)}
^{3}\Big]\omega_p ^{2} \\
& & \mbox{} - 3\,{({\imb p}\cdot{\imb k}_1)}^{4}\,{({\imb p}\cdot{\imb
    k}_2)}^{6}\,({ ({\imb p}\cdot{\imb k}_2)} + {({\imb p}\cdot{\imb
    k}_1)})\,(3\,{({\imb p}\cdot{\imb k}_1)}
+ {({\imb p}\cdot{\imb k}_2)})\\
&&+\quad(k_1^o,{\imb k}_1)\leftrightarrow(k_2^o,{\imb k}_2) \; ,
\end{eqnarray*}

\begin{eqnarray*}
\lefteqn{B(K_1,K_2)={k_1^o}\,{k_2^o}^{2}\,\Big[{k_1^o}\,{({\imb
      p}\cdot{\imb k}_2)}^{2}
  + {k_2^o}\,{({\imb p}\cdot{\imb k}_1)}^{2}\Big]\,\omega_p^{4}}\\
&& - {k_2^o}\,{({\imb p}\cdot{\imb k}_1)}^{2}\, \Big[({k_1^o}+
{k_2^o})\,{({\imb p}\cdot{\imb k}_2)}^{2} - {k_2^o}\,{({\imb
    p}\cdot{\imb k}_1)}
({\imb p}\cdot({\imb k}_1+{\imb k}_2))\Big]\omega_p^2\\
& & \mbox{} - {({\imb p}\cdot{\imb k}_1)}^{2}\,{({\imb p}\cdot{\imb
    k}_2)}^{3}\,({\imb p}\cdot({\imb k}_1
+{\imb k}_2))\\
&&+\quad (k_1^o,{\imb k}_1)\leftrightarrow(k_2^o,{\imb k}_2) \; .
\end{eqnarray*}

\section{Calculation of $
  \int\nolimits_{0}^{+\infty} x^n \ln(x)(\exp(x)+1)^{-p} dx$}
\label{app:integrals}

In the section \ref{sec:Pis}, we need to evaluate integrals of the
form:
\begin{equation}
I_{n,p}\equiv\int\limits_{0}^{+\infty}dx\,{{x^n\,\ln(x)}
\over{\left(e^x+1\right)^p}}
\; ,
\end{equation}
where $n,p$ are positive integers.  The starting point is to expand
$(e^x+1)^{-p}$ in powers of $e^{-x}$, which gives:
\begin{equation}
I^1_{n,p}={{(-1)^p}\over{(p-1)!}}\sum\limits_{m=1}^{+\infty}(-1)^m 
\sum\limits_{i=0}^{p-1}\alpha_{p-1,i}\; m^i \int\limits_{0}^{+\infty}
dx\, x^n\, \ln(x)\,e^{-mx}\; ,
\end{equation}
where the numbers $\alpha_{p-1,i}$ are the coefficients of the
polynomial
\begin{equation}
Q_{p-1}(x)\equiv (x-1)(x-2)\cdots(x-p+1)\equiv
\sum_{i=0}^{p-1}\alpha_{p-1,i}\,x^i\; .
\end{equation}
We need then
\begin{equation}
\int\limits_{0}^{+\infty}
dx\, x^n\, \ln(x)\,e^{-mx}={{A_n-n!(\gamma+\ln(m))}\over{m^{n+1}}}\; ,
\end{equation}
where $\gamma$ is the Euler's constant, and $A_n$ are integers defined
recursively\footnote{The solution of the recursion is $A_n=\gamma
  n!+\Gamma^\prime(n+1)$ where
  $\Gamma(z)=\int\nolimits_{0}^{+\infty}\!dt\,e^{-t}t^{z-1}$, but this
  expression does not make obvious the fact that $A_n$ is an integer.}
by
\begin{equation}
A_0=0\qquad
A_{n+1}=n!+(n+1)A_n\; .
\end{equation}
It is now straightforward to collect the various pieces in order to
obtain the following expression
\begin{eqnarray}
  &&\!\!\!\!\!I_{n,p}={{(-1)^pn!}
    \over{(p-1)!}}\sum\limits_{i=0}^{p-1}\alpha_{p-1,i} \left[
    \vphantom{+(2^{i-n}-1)\left({{A_n}\over{n!}}
        -\gamma\right)\zeta(n+1-i)} (2^{i-n}\!-\!1)\zeta^\prime(n+1-i)
    -2^{i-n}\ln(2)\zeta(n+1-i)
  \right.\nonumber\\
  &&\qquad\qquad\qquad\qquad\qquad\left
    .+(2^{i-n}-1)\left({{A_n}\over{n!}}  -\gamma\right)\zeta(n+1-i)
  \right]\; .
\label{eq:ln-integral}
\end{eqnarray}

\end{document}